\newcommand{\PRL}[3]{Phys.\ Rev.\ Lett.\ {\bf #1},\ #2 (#3)}
\newcommand{\RMP}[3]{Rev.\ Mod.\ Phys.\ {\bf #1},\ #2 (#3)}

\newcommand{\NAT}[3]{Nature\ {\bf #1},\ #2 (#3)}
\newcommand{\SC}[3]{Science\ {\bf #1},\ #2 (#3)}
\newcommand{\NATPHYS}[3]{Nature Phys.\ {\bf #1},\ #2 (#3)}

\newcommand{\PRA}[3]{Phys.\ Rev.\ A\ {\bf #1},\ #2 (#3)}
\newcommand{\PRB}[3]{Phys.\ Rev.\ B\ {\bf #1},\ #2 (#3)}

\newcommand{\JPC}[3]{J.\ Phys.:\ Cond.\ Mat.\ {\bf #1},\ #2 (#3)}

\newcommand{\PHYSREV}[3]{Phys.\ Rev.\ {\bf #1},\ #2 (#3)}

\newcommand{\diracslash}[1]{#1\llap{/\kern2pt}}

\newcommand{\be}{\begin{equation}}
\newcommand{\ee}{\end{equation}}
\newcommand{\bea}{\begin{eqnarray}}
\newcommand{\eea}{\end{eqnarray}}
\newcommand{\ba}[1]{\begin{array}{#1}}
\newcommand{\ea}{\end{array}}

\documentclass[twocolumn,superscriptaddress,secnumarabic,amssymb,nobibnotes,nofootinbib,aps,pra,showpacs]{revtex4-1}
\usepackage{graphicx}
\begin{document}

\title{Effect of Disorder in BCS-BEC Crossover}

\author{Ayan Khan}\thanks{ayan.khan@pusan.ac.kr}
 \affiliation{ Research Center for Dielectric and Advanced Matter Physics, Pusan National University, Busan, 609-735, S. Korea}
\author{Saurabh Basu}
\affiliation{Department of Physics, Indian Institute of Technology Guwahati, Guwahati, Assam, 781039, India}
 \author{Sang Wook Kim}%\thanks{swkim0412@pusan.ac.kr}
 \affiliation{Department of Physics Education, Pusan National University, Busan, 609-735, S. Korea}

\date{\today}

\begin{abstract}
%A systematic study of BCS-BEC crossover in presence of tender random impurity disorder is presented here. 
%The disorder in been included through the Nozi\`{e}res and Smith-Rink (NSR) theory of superconducting fluctuations. 
%Hence forth we 
%show that the condensate fraction shows a non monotonic behavior near the unitarity suggesting a more robust 
%paradigm of superfluidity. Further we calculate the collective excitation mode, which suggest
%a decrease in the sound velocity for composite bosons.
In this article we have investigated the effect of
weak random disorder in the BCS-BEC crossover region. The disorder is
included in the mean field formalism through NSR theory of superconducting
fluctuations. A self consistent numerical solution of the coupled equations involving  
the superfluid gap parameter and density as a function of the disorder strength, albeit 
unaffected in the BCS phase, yields a depleted order parameter in the BEC regime and an interesting  
nonmonotonic behavior of the condensate fraction in the vicinity  of  the unitary region, and a gradual 
depletion thereafter, as the pairing interaction is continuously tuned across the BCS-BEC crossover. 
The unitary regime thus demonstrates a robust paradigm of superfluidity even when the disorder is introduced.
To support the above feature and 
shed light on a lingering controversial issue, we have computed the behavior of the sound mode across the 
crossover that distinctly reveals a suppression of the sound velocity. 
We also find the Landau critical velocity that shows 
similar nonmonotonicity as that of the condensate fraction data, thereby supporting a stable superfluid scenario 
in the unitary limit.
\end{abstract}


\maketitle
\section{Introduction} 
Atomic gases at very low temperature are a unique system where one can observe the continuous evolution of a fermionic system
(BCS type) to a bosonic system (BEC type) 
by changing the inter-atomic interaction by means of Fano-Feshbach resonance \cite{grimm}. The experimental
advances to address this transition has introduced the possibility for
studying the so called BCS-BEC crossover more closely \cite{stringari1,dalibard1}. The physics of the crossover 
focuses on the change of s-wave scattering length ($1/a$) from attractive to repulsive (from $-\infty$ to $+\infty$) 
by tuning an external magnetic field. In the intermediate region where $1/a\rightarrow 0$: a new region
emerges, where dimensionless coherence length becomes in the order of unity, 
%and normalized critical temperature becomes 
%similar to the to the high temperature superconductors, 
is the focal point of crossover physics. 

The situation becomes further interesting if one assumes existence of random disorder in an otherwise very 
clean system. It is well-known that every wavefunction is spatially localized when the disorder is introduced 
in one-dimensional case, which is called as Anderson localization \cite{anderson1}.
However, it is  immensely difficult to directly
observe the localization in electronic systems on crystal lattices, 
so that one has to take the indirect route 
of conductivity measurement to observe the effects of localization. 
It is an intriguing question how the Anderson localization modifies the BCS superconductivity. 
It has been found by Anderson \cite{anderson2} that the order parameter is unaffected when the disorder is not 
too strong to give rise to the Anderson localization. Here the time reversal pairs, rather than the 
opposite momentum pairs, then form the Cooper pairs.

The cold atomic system offers great amount of controllability and allows one to observe macroscopic wavefunction. Therefore,
it was conceived as a very useful candidate to visualize the disorder effects directly. Recently the ultracold  
Bose gas ($^{87}Rb$ and $^{39}K$) enabled us to see the localization directly \cite{billy1,roati1}. 
Latest experiment are conducted in three dimension for both noninteracting atomic Fermi gas of 
$^{40}K$ \cite{kondov} and bose gas of $^{87}Rb$ \cite{bouyer}. 
These experiments has widened the possibility to study the crossover in lights of disorder 
\cite{lewenstein1} experimentally.

The static disorder in Fermi and Bose systems are not new issues.
A considerable amount of attention has been paid to disorder in superconductors \cite{belitz,trivedi1,trivedi2} 
and in Bose gas 
\cite{pavloff1,gaul,lopatin,lewenstein2,lewenstein3,pavloff2,palencia1,palencia2,palencia3,pilati,modugno,shapiro}.
Of late the interest at unitarity is also gaining pace \cite{orso,shklovskii,sademelo1,basu,sademelo2,sademelo3}, 
but still it failed to address all the questions associated with it. It is exciting to envision the three dimensional 
phase diagram
involving temperature, interaction and disorder through the evolution from BCS to BEC superfluid.  At the beginning it can be considered that the random potentials are independent of the 
hyperfine states of the atoms but then it can be extended to the correlated disorder problem in the crossover region.

In this article, we present our investigation on BCS-BEC crossover with weak uncorrelated disorder at zero temperature. 
Precisely, we show: (i) monotonic depletion of order parameter,
(ii) nonmonotonic nature of condensate fraction, and 
(iii) suppression of sound velocity, 
as a function of disorder. Further we add a study on Landau critical velocity 
to attempt a qualitative understanding of the nonmonotonic behavior of the condensate fraction. 
Though order parameter is mainly for academic interest but the other three quantities are experimentally 
viable \cite{ketterle1,ketterle4,ketterle3}.
In this study, we observe that superfluidity is more robust in the unitary regime, which is also consistent with behavior 
of the coherence length in the crossover regime. 
The robust nature of superfluidity has already been pointed out in the context of vortex core structure \cite{sensharma},
Josephson current \cite{strinati1} and collective modes \cite{stringari2}. 
We also observe a progressive decay of the sound velocity in the BEC side with increasing disorder possibily occuring  
due to enhancement of impurity scattering \cite{yukalov2}. 
Here we include a systematic study of the 
physical observables and their response to the random disorder.

The recent experiments on ultracold Bose and Fermi gas with disorder was carried out with optical 
speckle \cite{billy1,kondov}
and quasi periodic optical lattices \cite{roati1}. Though they pose interesting physics to study, but here we
consider quenched delta correlated disorder which remained subject of interest at zero temperature \cite{orso} and at finite 
temperatures \cite{sademelo3} in previous studies. 
One can visualize the situation of impurity driven random potential by considering the presence of few heavy 
atoms (say $^{40}K$) in a homogeneous bath of light $^6Li$ atoms \cite{sademelo3}. Since the number of heavy atoms are very 
limited, one can term it as a quasi-homogeneous system. 
To include the disorder effects in the mean field approach, we follow the NSR theory 
\cite{nsr} of superconducting fluctuations extended to broken symmetry state \cite{sademelo4}. Further we take advantage 
of the techniques developed to study condensate fraction of clean Fermi gas with Gaussian fluctuations \cite{taylor1,taylor2}.

We arrange our study in the following way, in Section II we present the basic formalism. Section III is dedicated to discuss 
our results in three parts. First part contains the order parameter, second one for the condensate fraction 
and the third leads to sound mode and Landau critical velocity. Finally, in Section IV we draw our conclusions.
\section{Formalism}
To make our presentation self-contained, we briefly summarize the mathematical formalism presented in Ref.\cite{orso}.
To describe the effect of impurity in Fermi superfluid in the crossover from BCS to BEC regime one needs to start from the real space 
Hamiltonian in three dimension for a $s$-wave superfluid,
\begin{eqnarray}
 \mathcal{H}(\mathbf{x})&=&\sum_{\sigma}\Phi^{\dagger}_{\sigma}(\mathbf{x})\Big[-\frac{\nabla^2}{2m}-
\mu+\mathcal{V}_{d}(\mathbf{x})\Big]\Phi_{\sigma}(\mathbf{x})+\nonumber\\
&&\int dx'\mathcal{V}(\mathbf{x},\mathbf{x'})\Phi_{\uparrow}^{\dagger}(\mathbf{x'})
\Phi_{\downarrow}^{\dagger}(\mathbf{x})\Phi_{\downarrow}(\mathbf{x})\Phi_{\uparrow}(\mathbf{x'}),\label{H}
\end{eqnarray}
where $\Phi_{\sigma}^{\dagger}(\mathbf{x})$ and $\Phi_{\sigma}(\mathbf{x})$ represents the creation 
and annihilation of fermions with mass $m$ and spin state $\sigma$
at $\mathbf{x}$ respectively. $\mathcal{V}_{d}(\mathbf{x})$ signifies the (weak) random potential and $\mu$ is the 
chemical potential. We set $\hbar =1$, where $\hbar$ is the Planck constant.
The s-wave fermionic interaction is defined as
$\mathcal{V}(\mathbf{x},\mathbf{x'})=-g\delta(\mathbf{x}-\mathbf{x'})$. The disorder potential is modeled as,
$\mathcal{V}_{d}(\mathbf{x})=\sum_{i}g_{d}\delta(\mathbf{x}-\mathbf{x}_{i})$ where $g_{d}$ is fermion-impurity coupling 
constant and $\mathbf{x}_{i}$ are the static positions of the quenched disorder. We assume it exhibits white noise 
correlation, that is,
$\langle \mathcal{V}_{d}(-q)\mathcal{V}_{d}(q)\rangle=\kappa\beta\delta_{i\nu_{m},0}$. $\beta$ is the inverse 
temperature, $\nu_{m}$ is the 
bosonic Matsubara frequency and $\kappa=n_{i}g_{d}^2$, that describes the strength of the impurity potential with
$n_{i}$ being the concentration of the impurities. 
%For unit volume,
%$v_{d}\ll\epsilon_{F}$; and if range of the impurity potential is $l_{d}$ then $n_{i}\ll1/l_{d}^3$.

The partition function corresponding to Eq.(\ref{H}) can be written in the path integral formulation as 
\begin{eqnarray}
\mathcal{Z}=\int\mathcal{D}[\bar{\Phi},\Phi]\exp{[-\mathcal{S}(\{\bar{\Phi}\}\{\Phi\})]},\label{newz} 
\end{eqnarray}
where
$\mathcal{S}=\int_{0}^{\beta}d\tau\int d\mathbf{x}[\bar{\Phi}_{\sigma}\partial_{\tau}\Phi_{\sigma}+\mathcal{H}]$.
By introducing the pairing field $\Delta(\mathbf{x},\tau)$ and by applying the Grassman identity
($\int\mathcal{D}[\bar{\Delta},\Delta]\exp{[-\int d\mathbf{x}\int_{0}^{\beta}d\tau \bar{\Delta}\Delta/g]}=1$) 
Eq.(\ref{newz}) can be given as
\begin{eqnarray}
 \mathcal{Z}&=&\int\mathcal{D}[\bar{\Phi},\Phi]\int\mathcal{D}[\bar{\Delta},\Delta]\exp{[-\mathcal{S}_{eff}]},\label{Z}
\end{eqnarray}
where $\mathcal{S}_{eff}=\mathcal{S}({\bar{\Phi},\Phi})+1/g\int d\mathbf{x}\int_{0}^{\beta}d\tau\bar{\Delta}\Delta$.
Following Hubbard-Stratonovich transformation, Eq.(\ref{Z}) can be written in terms of inverse Nambu propagator as,
\begin{eqnarray}
 \mathcal{Z}_{eff}&=&\int\mathcal{D}[\bar{\Delta},\Delta]e^{-1/g\int d\mathbf{x}\int_{0}^{\beta}d\tau\bar{\Delta}\Delta}
\nonumber\\
&\times&\int\mathcal{D}[\bar{\Phi}\Phi]e^{-\int d\mathbf{x}\int_{0}^{\beta}d\tau\bar{\Phi}\mathcal{G}^{-1}\Phi},\label{zeff}
\end{eqnarray}
where the inverse Nambu propagator $\mathcal{G}^{-1}(\mathbf{x},\tau)$ is defined as,
\begin{eqnarray}                               
\left(\begin{array}{ll}
-\partial_{\tau}+\frac{\nabla^2}{2m}+\mu-\mathcal{V}_{d} & \qquad\quad\Delta\\
\qquad\quad\bar{\Delta} & -\partial_{\tau}-\frac{\nabla^2}{2m}-\mu+\mathcal{V}_{d}
           \end{array}\right).\label{G1}
\end{eqnarray}
After integrating out the fermionic fields from Eq.(\ref{zeff}) we are left with the effective action as,
\begin{eqnarray}
 \mathcal{S}_{eff}&=&\int d\mathbf{x}\int_{0}^{\beta}d\tau
\Big[\frac{|\Delta(\mathbf{r})|}{g}-\frac{1}{\beta}\mathrm{Tr}\ln\{-\beta \mathcal{G}^{-1}(\mathbf{r})\}\Big],\label{seef}
\end{eqnarray}
where $\mathbf{r}=(\mathbf{x},\tau)$.
It is important to mention that the main contribution in the partition function comes from a small fluctuation,
$\delta\Delta(\mathbf{x},\tau)=\Delta(\mathbf{x},\tau)-\Delta$ where $\Delta$ is the homogeneous BCS pairing field.
The Green's function in Eq.(\ref{G1}) can be written as a sum of Green's function in absence of disorder 
($\mathcal{G}_{0}^{-1}=-\partial_{\tau}\mathbb{I}+(\nabla^2/2m+\mu)\mathbb{\sigma}_{z}+\Delta\mathbb{\sigma}_{x}$) 
and a self energy contribution 
($\Sigma=-\mathcal{V}_{d}\mathbb{\sigma}_{z}+\delta\Delta\mathbb{\sigma}_{+}+\bar{\delta\Delta}\mathbb{\sigma}_{-}$) 
which contains the disorder as well as the small fluctuations of the BCS pairing fields. $\mathbb{I}$ denotes the identity 
matrix, and $\sigma_{i}$ are the Pauli matrices and ladder matrices($i\in \{x,y,z,+,-\}$). 

By expanding the inverse Nambu propagator upto the second order one can write the effective action ($S_{eff}$)
in Eq.(\ref{seef}) as a sum of bosonic action ($S_{B}$) and fermionic action ($S_{F}$). Also it contains an 
additional term which emerges from the linear order of self 
energy expansion ($G_{0}\Sigma$). It is possible to set zero for the linear order if we consider 
$S_{F}$ is an extremum of $S_{eff}$ after performing all the fermionic 
Matsubara frequency sums. The constrained condition leads to the BCS gap equation which after appropriate regularization 
through the $s$-wave scattering length reads,
\begin{eqnarray}
 -\frac{m}{4\pi a}=\sum_{k}\Big[\frac{1}{2E_{k}}-\frac{1}{2\epsilon_{k}}\Big].\label{gap}
\end{eqnarray}
Eq.(\ref{gap}) suggests that the BCS gap equation does not have any contribution from the disorder potential explicitly. 

Now to construct the 
density equation with usual prescription of statistical mechanics; the thermodynamic potential $\Omega$ should be
differentiated with respect to the chemical potential $\mu$. $\Omega$ can be written as a sum over fermionic ($\Omega_{F}$)
and bosonic ($\Omega_{B}$) thermodynamic potentials, which implies, 
\begin{eqnarray}
 n=n_{F}+n_{B}=-\frac{\partial}{\partial\mu}(\Omega_{F}+\Omega_{B})=
-\frac{1}{\beta}\frac{\partial}{\partial\mu}(S_{F}+S_{B}).\label{nfnb}
\end{eqnarray}
%$n=n_{F}+n_{B}=-\frac{\partial}{\partial\mu}(\Omega_{F}+\Omega_{B})=
%-\frac{1}{\beta}\frac{\partial}{\partial\mu}(S_{F}+S_{B})$. 
The well known BCS density equation can be restored from Eq.(\ref{nfnb})
if we consider only $n_{F}$, then it yields the familiar $\sum_{k}(1-\frac{\xi_{k}}{E_{k}})$. However the presence of disorder 
and
fluctuation leads to $n_{B}\neq0$. Hence the final mean field density equation will be,
\begin{eqnarray}
 n&=&\sum_{k}\Big(1-\frac{\xi_{k}}{E_{k}}\Big)-\frac{\partial\Omega_{B}}{\partial\mu}.\label{density}
\end{eqnarray}
The bosonic thermodynamic potential consists of two parts. One comes from the thermal contribution 
and the other is due to disorder. Since we are interested in zero temperature  
we neglect the thermal contribution from here on. Henceforth the disorder induced thermodynamic potential
can be written as,
\begin{eqnarray}
 \Omega_{B_{d}}&=&-\frac{\kappa}{2}\sum_{\mathbf{q},\nu_{m}=0}\mathcal{N}^{\dagger}\mathcal{M}^{-1}\mathcal{N}.\label{obd}
\end{eqnarray}
%$-\frac{\kappa}{2}\sum_{\mathbf{q},\nu_{m}=0}\mathcal{N}^{\dagger}\mathcal{M}^{-1}\mathcal{N}$.
%where $\Omega_{B}$ is defined as,
%\begin{eqnarray}
% \Omega_{B}&=&\lim_{\beta\rightarrow\infty}\frac{1}{2\beta}\sum_{q}\ln\det \mathcal{M} -
%\frac{\kappa}{2}\sum_{\mathbf{q},\nu_{m}=0}\mathcal{N}^{\dagger}\mathcal{M}^{-1}\mathcal{N}.\label{omegab}
%\end{eqnarray}
%The first term in Eq.(\ref{omegab}) is the contribution of the fluctuating pairing fields.
%This contribution becomes important if the objective of study
%relates to the finite temperature effect on the system.
%Since our purpose of study is the effect of disorder at zero temperature, from here on we will neglect this
%contribution. In Eq.(\ref{omegab}) the $\mathcal{N}$ is a doublet which couples the disorder with the fluctuation. 
where $\mathcal{N}$ is a doublet which couples disorder with fluctuation.
After 
performing the fermionic Matsubara frequency summation over $\mathcal{N}$,
\begin{eqnarray}
 \mathcal{N}_{1}=\mathcal{N}_{2}=\sum_{k}\frac{\Delta(\xi_{k}+\xi_{k+q})}{2E_{k}E_{k+q}(E_{k}+E_{k+q})}.
\end{eqnarray}
The inverse fluctuation propagator matrix $\mathcal{M}$ is a $2\times2$ symmetric matrix whose elements are given by
\begin{eqnarray}
 \mathcal{M}_{11}&=&\frac{1}{g}+\sum_{k}\biggl[\frac{v_{k}^{2}v_{k+q}^{2}}{i\nu_{m}-E_{k}-E_{k+q}}-
\frac{u_{k}^{2}u_{k+q}}{i\nu_{m}+E_{k}+E_{k+q}}\biggr],\nonumber\\
\mathcal{M}_{12}&=&\sum_{k}u_{k}v_{k}u_{k+q}v_{k+q}\biggl[\frac{1}{i\nu_{m}+E_{k}+E_{k+q}}\nonumber\\
&-&\frac{1}{i\nu_{m}-E_{k}-E_{k+q}}\biggr],
\end{eqnarray}
and $\mathcal{M}_{22}(q)=\mathcal{M}_{11}(-q)$. Here the usual BCS notations have been used, namely 
$\xi_{k}=\mathbf{k}^2/2m-\mu$, $E_{k}=\sqrt{\xi_{k}^2+\Delta^2}$, $u_{k}^2=\frac{1}{2}(1+\xi_{k}/E_{k})$ and 
$v_{k}^2=\frac{1}{2}(1-\xi_{k}/E_{k})$.
\section{Results}
\subsection*{Order Parameter}
Eqs.(\ref{gap}) and (\ref{density}) are now ready to be solved self consistently. 
Our analysis is valid only for the weak disorder. Considering Ref.\cite{sademelo3} 
we can safely assume that the disorder is weak 
if the dimensionless disorder strength $\eta(=\kappa m^2/k_{F}) \lesssim 5$ is satisfied.
%It is clear that 
%the disorder strength, $\kappa$ is an input parameter. This provokes the question about how to estimate 
%the disorder strength, that is, distinguish the case of a weak disorder from that of a strong one. 
%Firstly, $\kappa$ has a dimension of $k_{F}/m^2$. So in our analysis we define a dimensionless
%disorder strength $\eta=\kappa m^2/k_{F}$. But a more physical description can be worked out if the impurity
%strength is normalized by Fermi density and square of Fermi energy \cite{sademelo3}. According to the this new definition, 
%$\kappa n_{F}/\epsilon_{F}^2=4/(3\pi^2)\eta$. So one can remain in the weak disorder limit 
%as long as the impurity coupling constant $g_{d}<<\epsilon_{F}$ and the concentration or density of the impurity 
%$n_{i}<<n_{F}$. 
%It is already argued that one can remain in the weak disorder
%limit as long as range of the impurity potential is much smaller than the mean free path of the scatterers. 
%In this language
%if $l_{F}$ is the elastic mean free path of of unbound fermions then $k_{F}l_{F}>>4/(3\pi\tilde{\eta})$ \cite{sademelo3}.
%Thus we calculate self consistently the basic mean field quantities $\Delta$ and $\mu$ being inside this limit for different
%values of $\eta$.
\begin{figure}
 \begin{center}
  \includegraphics[width=9cm]{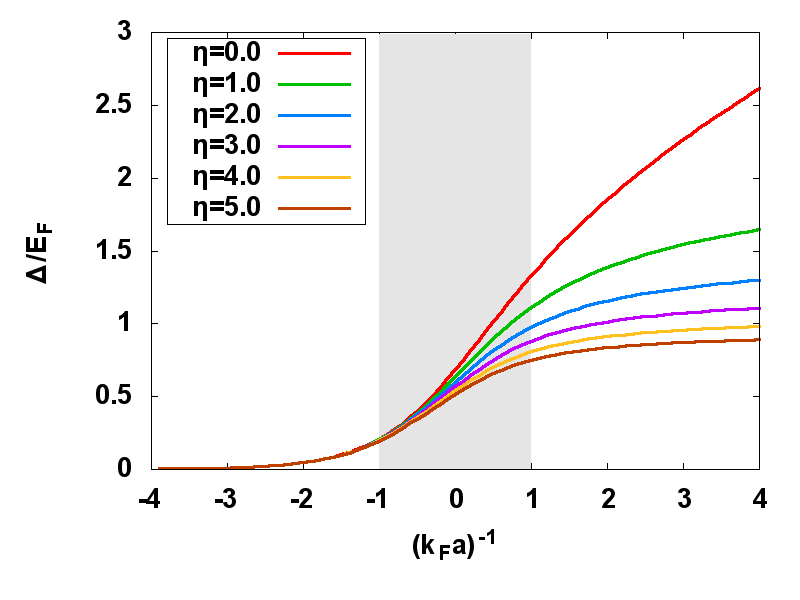}
\caption{The order parameters $\Delta$ normalized by Fermi energy as function of $(k_{F}a)^{-1}$ for 
various disorder strengths. 
The shaded area represents the crossover region}\label{gap_plot}
 \end{center}
\end{figure}
Fig. \ref{gap_plot} demonstrates that in the BCS limit ($1/k_Fa -> -\infty$) every order parameters with 
different disorder strength follows the mean field expression of 
$\Delta/\epsilon_{F}=8e^{-2}\exp[-\pi/(2k_{F}a)]$  
emphasizing the validity of Anderson theorem \cite{anderson2}. In the BEC limit, instead,
one can observe a progressive depletion. 
Assymptotically this depletion  
remains in the order of $\eta/(k_{F}a)$ \cite{orso,huang}.
The other mean field quantity, chemical potential, does not change much(not shown). 
We understand that it might be an attribute of the fluctuation theory \cite{sensharma} where the correction in 
$\mu$ at the BCS side is usually in $\mathcal{O}(\Delta^2)$ which is a very small quantity
as $\Delta\rightarrow0$ for $(k_{F}a)^{-1}\rightarrow0$. In the BEC side
the correction comes through the effective chemical potential of the composite bosons. However the BEC chemical
potential is dominated by the binding energy and it turns out quite large compared to the effective chemical potential.
The external potential $\mathcal{V}_{d}(x)$ induced by the disorder usually has no direct influence on the internal degrees of freedom, 
e.g. interaction among fermions. It is 
thus no wonder that the binding energy of the composite bosons exhibits no pronounced change, so does the chemical potential.
%In our calculation we have also observed very small deviation from the clean limit, but as described they
%are not pronouced enough to present.
%The disorder stimulated the superfluid to transform into normal fluid but the total number of particles are conserved hence
%one can not see much change in the chemical potential. 
\begin{figure}
 \begin{center}
  \includegraphics[width=9cm]{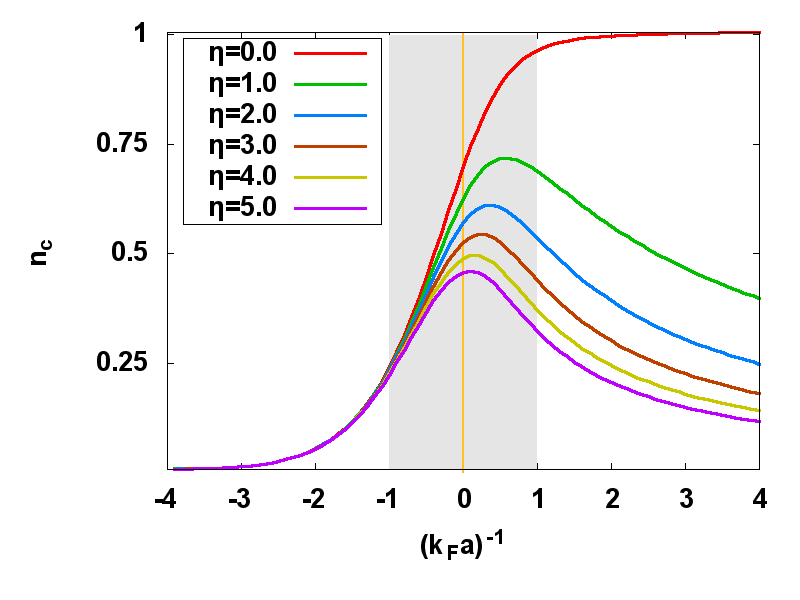}
\caption{The condensate fraction $n_{c}$ as function of $(k_{F}a)^{-1}$ for various disorder strengths. 
The shaded area and the vertical orange line at 
$(k_{F}a)^{-1}=0$ represent the crossover region and the unitary limit, respectively.}\label{cfrac_plot}
 \end{center}
\end{figure}
\subsection*{Condensate Fraction}
Though $\Delta$ and $\mu$ are the first quantities to study the crossover through the mean field theory but they are mostly
of academic interest. From now on we shall focus on the quantities which are more prone to the experimental observation.
Our first choice is the condensate fraction which is also one of our main results (depicted in Fig.\ref{cfrac_plot}).
In a clean Fermi gas, it is possible to work out the condensate fraction through mean field theory \cite{salasnich} which
shows good agreement with the experiment \cite{ketterle4}. Here we follow the similar mean field description,
\begin{eqnarray}
n_{c}=\sum_{k}\Big[\frac{\Delta(\eta)}{2E_{k}(\eta)}\Big]^2, \label{cfrac}
\end{eqnarray}
where $n_{c}$ is the condensate fraction.
The only difference with the clean system calculation, in Eq.(\ref{cfrac}) is that, we used the 
disorder induced values for $\Delta$ and $\mu$.
We observe quite remarkable behavior for $n_{c}$. As one expects the condensate fraction decays 
similar to that of the clean limit  when $1/(k_{F}a)\rightarrow-\infty$ 
since $\Delta$ does not change with the variation of $\eta$ in the BCS regime. If one extends the self energy upto the
second order in the condensate fraction calculation one can as well observe the effect of disorder in the BCS limit, where
$n_{c}-n_{c_{0}}$ gets saturated at some finite value instead of exponential decay \cite{orso}, with $n_{c_{0}}$ denoting
the condensate fraction in the clean limit.
In the BEC side, the disorder destroys part of the condensate and turns it into a normal 
fluid. The condensate fraction approaches roughly $\eta/\sqrt{k_{F}a}$ 
as obtained from the study of hard sphere Bose gas in random disorder \cite{huang}. 

The nonmonotonic behavior in the 
crossover region (grey area in Fig. \ref{cfrac_plot}) is the most intriguing point. In the study of quantized 
vortices, one sees
that accumulation of a number of vortices becomes maximum in the crossover region \cite{ketterle2}. Later on, in a 
theoretical study of a single vortex it was also observed that the circulation current is maximum in the crossover regime
\cite{sensharma}. Similar feature was also reported in the Josephson current study \cite{strinati1}. This unique 
behavior was qualitatively attributed to the maximization of Landau critical velocity in this region. 
In general all these observations
actually points to the robustness of the superfluidity at the unitarity. But normal mean field does not show any precise 
nonmonotonic behavior for the condensate fraction. 
Here with a weak disorder, we are able to generate a picture which affirms the belief that 
in presence of low amount of impurity the condensate fraction is less affected across the crossover, 
implying a comparatively high yielding of superfluidity in this region. 

One should also take a note of the position of the extrema.
With the increase in disorder, we observe that the extrema slowly move towards $1/a\rightarrow0_{+}$ from $+\infty$, 
but never cross it as we are exhausted with weak impurity limit. Further increase of disorder will break down 
the weakness condition. This result agrees qualitatively with Ref.\cite{orso,sademelo3}, however in those cases
it has been suggested that the most robust region of superfluidity emerges when $1/a\rightarrow0_{-}$.
But there is no good justification for that. 
Here it is pointed, this behavior is consistent with that of the critical velocity discussed in the following section. 
\subsection*{Sound Mode and Critical Velocity}
Another important and experimentally relevant \cite{ketterle3} quantity is the lowest energy collective excitation of the
condensate. The nature of sound velocity, in presence of disorder, in a Bose gas has been studied quite extensively,
but it lacks a real consensus so that considerable
amount of ambiguity still exists. Using a perturbative  method, the sound velocity is enhanced in the 
presence of uncorrelated disorder 
and very weak interaction \cite{lopatin,stringari3,graham1,graham3}, whereas within a non-perturbative self 
consistent approach and spatially correlated weak disorder case, depression in sound velocity has 
been reported \cite{graham2,graham4,gaul1,yukalov}. In a more recent study it has been shown that there exists no
generic behavior of sound in the presence of disorder using a perturbative approach \cite{gaul2}. Hence in the crossover region,
the behavior of sound is expected to be quite interesting.
\begin{figure}
 \begin{center}
  \includegraphics[width=9cm]{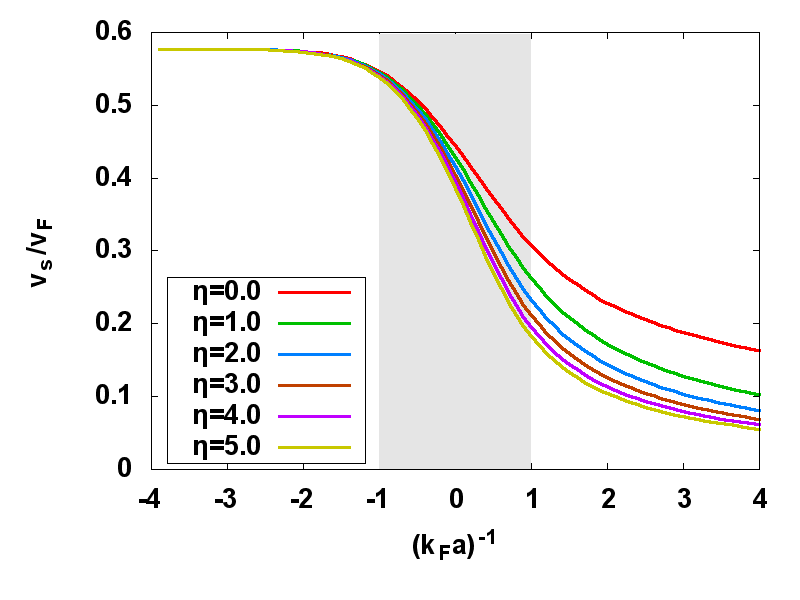}
\caption{The sound velocity $v_{s}$ divided by the Fermi velocity $v_{F}$ as a function of $(k_{F}a)^{-1}$ for 
various disorder strengths. 
The shaded area depicts crossover region.}\label{sound_plot}
 \end{center}
\end{figure}

From technical point of view in order to obtain the sound velocity %depicted in Fig (\ref{sound_plot}) 
one needs to carry out an analytic continuation of the Matsubara frequency.
%transfer the frequency from imaginary Matsubara axis to the real line
%($i\nu_{m}\rightarrow-\nu$). 
Hence the fluctuation propagator matrix $\mathcal{M}$ is expanded to the second order for both momentum 
and frequency. The determinant of which leads to \cite{strinati2},
\begin{eqnarray}
&& \mathcal{M}_{11}(\mathbf{q},\nu)\mathcal{M}_{12}(\mathbf{q},-\nu)-\mathcal{M}_{12}^{2}(\mathbf{q},\nu)=\nonumber\\
&&\mathcal{A}(\Delta,\mu,\mathbf{k})\mathbf{q}^2
+\mathcal{B}(\Delta,\mu,\mathbf{k})\nu^2+\dots=0,
\end{eqnarray}
for $\nu=v_{s}|\mathbf{q}|$ where $v_{s}$ represents the sound velocity.
$\mathcal{A}$ and $\mathcal{B}$ are functions of $\Delta$ and $\mu$ and can be evaluated by
summing them over $\mathbf{k}$. 
%The depletion in the sound mode in the BEC side may look erroneous from the preconceived 
%idea of regular Bose gas but a careful assessment shows that this must happen. 
In the BEC limit it has been shown that the sound velocity
is $v_{s}^2=\Delta^2/(8m|\mu|)$ \cite{stringari2} so that the suppression of the order parameter should 
directly result in that of the sound velocity. Near the unitarity, the sound velocity is directly connected to the
chemical potential through $v_{s}^2=2\mu/(3m)$. In effect, the sound velocities for various disorders are merged
near the unitarity as the chemical potential exhibits almost no change regardless of addition of the disorder.
Moreover, in the BCS side, $v_{s}=\sqrt{k_{F}^2/(3m^2)}\simeq0.57$ (in dimensionless units) 
does not have any explicit
dependence on the mean field parameters. Therefore, irrespective of the disorder the sound velocity saturates near $0.57$. 
All of these arguments successfully explain the behavior of the sound velocity shown in Fig. \ref{sound_plot}.

In an usual way, sound velocity is defined as $v_{s}=\sqrt{\frac{1}{\rho\kappa_{T}}}$ where $\rho$ is the mass density and 
$\kappa_{T}$ is the isothermal compressibility. Applying Gibbs-Duhem relation sound velocity can be written in usual notation
as $v_{s}=\sqrt{\frac{n}{m}\frac{\partial\mu}{\partial n}}$.
An involved study of sound mode from the thermodynamical point of view reveals that the decrease in the collective 
excitations
can be attributed to the increase in the compressibility. This is in accordance with Ref.\cite{graham2}. 
Intuitively this looks
more feasible as the occurrence of additional random potential should lead to additional scattering and,
resulting in the decrease of sound velocity \cite{yukalov2}.

Though the sound velocity itself is a quantity of huge interest, but it also serves an additional important information;   
According to Landau criterion, it determines
the critical velocity of BEC. The critical velocity is defined as the minimum velocity of the 
BEC required to break superfluidity by creating elementary excitations.
To obtain a clear picture leading to the critical velocity, we calculate the minimum velocity related to
the single particle
excitations induced by pair breaking, which is dominant in the BCS side.
%A direct calculation of the Landau critical velocity following the basic definition 
%$v_{c}=min\{E_{k}/{k}\}$, where $E_{k}$
%is the excitation energy with momentum $k$, one can see that in the BCS side the dominant contribution comes from the
%energy required for breaking a single pair or in other words from the single particle excitation spectrum. 
\begin{figure}
 \begin{center}
  \includegraphics[width=9cm]{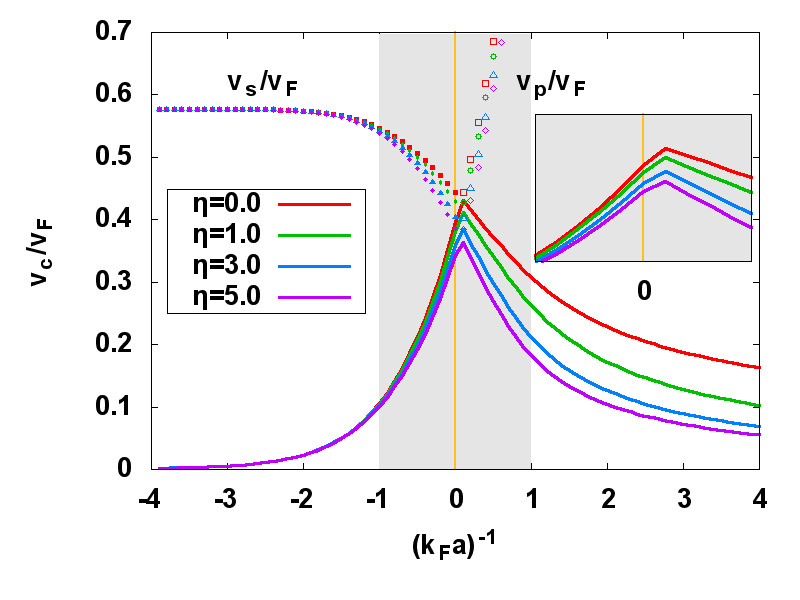}
\caption{The sound velocity $v_{s}$ (filled marks and solid curves) already shown in Fig. \ref{sound_plot} and 
the critical velocity 
$v_{p}$ (open marks and solid curves) obtained from pair breaking, normalized by the Fermi velocity $v_{F}$, as a function 
of $(k_{F}a)^{-1}$ 
for various disorder strengths. The critical velocity $v_{c}$'s (solid curves) are determined from $min\{v_{s},v_{p}\}$. 
The unitary region is depicted through 
the grey area. The inset focuses the turning point in the unitary region.}\label{lcrit_plot}
 \end{center}
\end{figure}
The fermionic single particle excitation within mean field 
is written as $\sqrt{(\sqrt{\Delta^2+\mu^2}-\mu)/m}$ \cite{stringari2}, which is reduced to the familiar result of 
$\Delta/k_{F}$ in the deep BCS region. In the case of the composite bosons, the Landau critical 
velocity is provided as the sound 
velocity. By choosing the minimum between $v_{s}$ and $v_{p}$, one can determine the Landau critical velocity, 
which is represented by the solid curves in Fig. \ref{lcrit_plot}.
%In Fig (\ref{lcrit_plot}), the solid lines corresponds to the Landau critical velocity whereas the opaque 
%points
%give the sound velocity ($v_{s}$) and transparent points are the pair breaking velocity $v_{p}$. 
%The colors codes are same for same disorder strengths. 

Though the simple mean field analysis on Landau critical velocity may not be
amenable for a direct mapping between $n_{c}$ and $v_{c}$, but it may provide a qualitative suggestion for less depletion of 
condensate fraction in the crossover regime. In both cases the maxima exist near the unitary 
regime and move toward $(k_{F}a)^{-1}=0$ as $\eta$ increases as shown in the inset of Fig. \ref{lcrit_plot}.
%In both the cases, the extrema exist near unitarity for 
%$1/a\rightarrow0_{+}$ and a slow movement of the peak towards the point of diverging scattering length can also be observed
%which is depicted in the inset of Fig (\ref{lcrit_plot}).

%The preceding discussion also enables us to estimate the phase coherence length or $\xi_{phase}$, which is associated 
%with the 
%spatial fluctuations of the superconducting (superfluid) order parameter. The phase coherence length, often 
%referred to as the healing length in the bosonic literature, is the length scale upto which the bulk condensate
%behavior is retained. As $\xi_{phase}\propto\frac{1}{v_{c}}$ \cite{stringari2}, hence from the previous analysis we can 
%easily make 
%out that at unitarity $\xi_{phase}$ is minimum. It is interesting to note that, the coherence length increases 
%in the vicinity of the diverging scattering length due to disorder, however the rate is not as severe as it is in the 
%case of bulk 
%composite bosons.
%In other words, the coherence length remains less affected to the random fields. This may lead to the nonmonotonic 
%behavior of the condensate fraction in the domain of the unitary regime. 

\section{Conclusion}
In conclusion, we have studied several important physical quantities such as gap parameter, condensate fraction and 
sound velocity etc. to address the issue of BCS to BEC crossover in a disordered environment. To this end
we have included weak disorder via the gaussian fluctuation as prescribed earlier 
\cite{orso,sademelo3}, and hence solved the coupled BCS mean field equations self consistently. This enables us to obtain 
the
two basic mean field parameters, $\Delta$ and $\mu$. We thus show that the order parameter gets depleted leaving
the chemical potential unchanged as we go from a BCS to a BEC regime. 
%This behavior was discussed in Ref.\cite{orso}. 
%Here we present calculations
%on these issues for the first time. 
Afterwards, the condensate fraction has been calculated using the well known mean 
field description
with the disorder affected $\Delta$ and $\mu$. As the disorder strength increases, 
we observe pronounced maximum developed in the crossover
region justifying the expectation of robust superfluid in this region. We have tried to connect this non monotonic nature 
qualitatively to the Landau critical velocity, which also shows a sharp maximum near unitarity. With increase of 
the impurity this peak slightly moves towards the unitary point. A similar feature is observed for the condensate fraction. 
In addition, the depression of the sound velocity is also been addressed, which might be related to the enhanced 
scattering from the random scatterers employed in this model. 

To be precise, the the nature of $\Delta$, $n_{c}$, $v_{s}$
and $v_{c}$ is been reported here, when subjected to a weak random impurity. 
We hope our present study has shed some lights on physics of the BCS-BEC crossover in disordered systems.
An interesting future perspective
can be generalization of the theory for arbitrarily strong disorder and interaction. 
We also hope that these results would be observed 
and verified in experiment soon.

\section*{Acknowledgement}
AK is grateful for useful communication with G. Orso and enlightening discussions with C. Sa de Melo, F. Dalfovo,
G. Roati, N. Trivedi and G. Watanabe. This was supported by the NRF grant funded by the Korea government (MEST) 
(No.2009-0087261 and No.2010-0024644).
SB acknowledges financial support from 
DST (SR/S2/CMP/0023/2009).

\end{document}